\documentclass[aps,pra,preprint,groupedaddress]{revtex4-1}
\usepackage{amsmath}
\usepackage{amsfonts}
\usepackage{amssymb}
\usepackage{graphicx}
\usepackage{epstopdf}
\usepackage[utf8]{inputenc}
\usepackage{caption}
\usepackage{subcaption}
\begin{document}
\title{Quantum vacuum emission from a refractive index front}
\author{Maxime Jacquet}
\author{Friedrich K\"{o}nig}
\email[]{fewk@st-andrews.ac.uk}
\affiliation{School of Physics and Astronomy, SUPA, University
of St. Andrews, North Haugh, St. Andrews, KY16 9SS, UK}
\date{\today}

\begin{abstract}
A moving boundary separating two otherwise homogeneous regions of a dielectric is known to emit radiation from the quantum vacuum.  An analytical framework based on the Hopfield model, describing a moving refractive index step in 1+1 dimensions for realistic dispersive media has been developed in \cite{finazzi2013}.  We expand the use of this model to calculate explicitly spectra of all modes of positive and negative norm.  Furthermore, for lower step heights we obtain a novel set of mode configurations.  This leads to a realistic emission spectrum featuring black hole- and white hole emission for different frequencies.  We also present spectra as measured in the laboratory frame that include all modes, in particular a dominant negative norm mode, which is the partner mode in any Hawking-type emission.  We find that the emission spectrum is highly structured into intervals of emission with black hole-, white hole-, and no horizons.  Finally we estimate the number of photons emitted as a function of the step height and find a power law of 2.5 for low step heights.
\end{abstract}

\maketitle

\section{Introduction}\label{sec:Introduction}

In an effort to tie the theory of quantum fields together with that of general relativity, Hawking showed  \cite{hawking1974,hawking1975} that black holes emit a steady thermal flux: in contradiction with their name, black holes are not black. The phenomenon, now called Hawking Radiation (HR), arises from quantum vacuum fluctuations at the event horizon.
It connects the realms of quantum physics, Einstein's theory of relativity, and thermodynamics. However, it is essentially impossible to observe in astrophysics because of its low temperature, which depends on the surface gravity at the horizon, i.e. the mass of the black hole.
A solar-mass black hole has a temperature well below that of the cosmic microwave background.

Fortunately, Unruh \cite{unruh1981} showed how wave media, if moving,  can be used to realize gravity analogues.
A flowing fluid exhibiting a gradient of flow velocity from subsonic to supersonic can mimic the flow of space at the event horizon: there is a boundary, beyond which an acoustic wave propagating against the flow will be carried downstream. This is the analogue of an event horizon.
In complete analogy, quantum hydrodynamical fluctuations in a moving fluid are thus predicted to convert into pairs of phonons at the sonic horizon.
Analogue systems produce analogue HR, providing a realistic chance to observe this particle emission from the quantum vacuum \cite{barcelo2005,barcelo2011}.

Until recently, proposed analogue systems to test Hawking's prediction by experiments suffered from a low characteristic temperature \cite{unruh1981,robertson2012,weinfurtner2011,garay2000,garay2001,nation2012,volovik2003, Nguyen2015}.
In 2008, Philbin et al \cite{philbin2008} demonstrated the feasibility of creating artificial event horizons with a moving refractive index front (RIF) in dispersive optical media.
They estimated a temperature of a thousand Kelvins and ushered-in the field of optical analogues \cite{robertson2011,leonhardt2009,belgiorno2010,finazzi2013}.
The idea behind optical event horizons is to create a change in the refractive index, i.e. the speed of light, with light itself.  For example, a pulse of light modifies the index by the optical Kerr effect. Light under the pulse will be slowed and thus the front of the pulse exhibits -for some frequencies- a black-hole type horizon capturing light. The back of the pulse acts as an impenetrable barrier, a white-hole horizon. 
The event horizon separates two discrete regions: under the pulse, where light is slow and the pulse moves superluminal, and outside the pulse, where the pulse speed is subluminal.

To date, optical HR is yet to be discovered. Detailed analytical predictions of the wavelength and intensity of the radiation for an actual experiment are difficult to make due to the extreme conditions at the horizon. In the above example, the length of the pulse front or back has to be comparable to the wavelength of radiation \cite{philbin2008}.
Nonetheless, progress has been made towards an understanding of the critical conditions needed to observe optical HR \cite{finazzi2012,finazzi2013,finazzi2014,KOENIG12,Petev2013,koenig2014,Robertson2014a,Robertson2014b, belgiorno2015}.
In particular, Finazzi and Carusotto \cite{finazzi2012,finazzi2013,finazzi2014,belgiorno2015} introduce a fully quantized analytical model for spontaneous emission at a moving boundary in a nonlinear dielectric. At this boundary between two multi-branch dispersive media certain modes may experience  either analogue black- or white hole or horizonless configurations, leading to mode mixing and spontaneous emission of radiation. 
All configurations lead to the spontaneous emission of radiation due to the mixing of modes with opposite norm.  In order to maintain these different configurations, they finely adapt the velocity of the RIF when changing the nonlinearity.  They discover that emission is dominant over optical frequencies.  Hence they focus on emission spectra in positive norm optical modes only.
All of quantum vacuum emission is accompanied by emission of negative norm modes, which play the role of the partner mode in the Hawking effect.  This is relevant, in particular, because these modes emit at different laboratory frequencies compared to their positive norm partner modes.  Furthermore, the nonlinearity in the experiment typically changes independently of the RIF velocity, leading to a very different spectral structure as well as scaling of the signal with nonlinearity/ intensity.
 
In the present paper we use the model \cite{finazzi2013} to reveal these properties of the quantum vacuum emission.  We first expand the model to consider emission from all positive and negative norm modes at any frequency and change of refractive index.  Hence we obtain emission spectra for different index changes without tuning of the RIF velocity.  This corresponds to most experimental situations where the velocity of a RIF would not change with height.  A resulting consequence is that the mode configurations can be more complicated than for a black- or a white hole.  In fact, we obtain spectra which combine black- and white hole emission. The system is then an object that simultaneously radiates as analogue black and white hole.  We convert the spectra to the laboratory frame, including all mode contributions, inclusive of the important negative norm ones.  The lab spectrum is characteristically structured into intervals of emission with black hole-, white hole-, and no horizons.  Finally, we observe that the total photon flux associated with black hole emission rises with a power of about 2.5 for low refractive index changes, whereas the emission bandwidth increases linearly.  For larger index changes,  $\delta n >0.05$, which are difficult to reach experimentally, the rise slows to a lower power law as the number of participating modes saturates.  These novel features will be useful to identify emission in future experiments on optical Hawking radiation.

We first introduce the theoretical model of the scattering of vacuum modes at the horizon.
We explain how the interaction of light and matter in a uniform dispersive medium is modeled.  We identify the eigenmodes and study their properties and then introduce the moving Refractive Index Front (RIF) separating sub and superluminal regions. We construct eigenmodes of the non-uniform medium with the RIF and describe the scattering/ mode conversion process at the RIF by the scattering matrix. We then quantize the field modes and calculate the photon flux density in the moving and laboratory frame.
Finally we compute the spectra of emission from bulk fused silica in both frames. We also integrate the spectra to evaluate the total emission and its scaling with the refractive index height. The spectra allow us to indentify in detail the contributions of the various modes to the emission and the role of analogue event horizons.

\section{Light matter interactions in a dispersive medium}\label{sec:Theory}

We start by laying out the canonical quantization scheme for the electromagnetic field in the linear and non-uniform dielectric. We restrict ourselves to a one-dimensional geometry and scalar electromagnetic fields and operate at frequencies sufficiently far from the medium resonances to neglect absorption.
Our analysis is based on the Hopfield model \cite{hopfield1958} and we follow the treatment presented in \cite{finazzi2013} before expanding it to include all modes at any frequency and refractive index change.

The step in refractive index (RIF) is propagating in the positive x-direction and the electric field is orthogonal in the z-direction.
Light propagation in the \textit{homogeneous} medium is sufficiently well described by the interaction with a set of three polarization fields, which phenomenologically reproduce the dispersion relation of transparent dielectrics. We state the Lagrangian density $\mathcal{L}$ describing the coupling of the vector potential $A(x,t) $ with the three polarization fields $P_i$, harmonic oscillators of elastic constant $\kappa_i^{-1}$ and inertia $\frac{\lambda_i^2}{\kappa_i (2 \pi c)^2}$, in the laboratory reference frame:
\begin{equation}
\label{eq:LagDens}
\mathcal{L}=\frac{(\partial_t A)^2}{8 \pi c^2}-\frac{(\partial_x A)^2}{8 \pi}+\sum_{i=1}^3{\left(\frac{(\partial_t P_i)^2\lambda_i^2}{2 \kappa_i (2 \pi c)^2}-\frac{P_i^2}{2 \kappa_i}+\frac{A}{c}\partial_t P_i\right)}.
\end{equation}
The term linear in \(A\) in Eq. (\ref{eq:LagDens}) describes the coupling between the fields.
This Lagrangian density accounts for the free space and medium contributions to the field through the first two terms and the sum, respectively.
Dispersion enters as a time dependence of the addends of the summation.

It is convenient to express the Lagrangian density in the moving frame by applying a Lorentz boost.
The system is stationary in the reference frame comoving with the RIF at velocity \(u\). We denote the space and time coordinates in the moving frame by, respectively, \(\zeta=\gamma\left(x-ut\right)\) and \(\tau=\gamma\left(t-x\frac{u}{c^2}\right)\) with the Lorentz factor \(\gamma=\left(1-u^2/c^2\right)^{-1/2}\).
We obtain the Hamiltonian density by varying the Lagrangian density with respect to the canonical momentum densities \(\Pi_A=\frac{\partial \mathcal{L}}{\partial(\partial_{\tau} A)}\) and \(\Pi_{P_i}=\frac{\partial \mathcal{L}}{\partial(\partial_{\tau} P_i)}\) of the light and polarization fields.
From there follow the Hamilton-Jacobi equations, the equations of motion for the fields.\par

The linear Lagrangian (\ref{eq:LagDens}) leads to the generic Sellmeier dispersion relation of bulk transparent dielectrics:
\begin{equation}
c^2 k^2=\omega^2 \left[1+\sum_{i=1}^3\frac{4\pi\kappa_i}{1-\frac{\omega^2\lambda_i^2}{(2 \pi c)^2}}\right]
\label{eq:LabSellmeier}
\end{equation}
where \(\omega\) and \(k\) are the laboratory frequency and wave number, respectively.
The dispersion relation (\ref{eq:LabSellmeier}) is plotted in fig. \ref{fig:LabDispRel}: there are eight branches, four with positive laboratory frequency and their four negative frequency counterparts - symmetric about the \(k\)-axis.
To find orthonormal plane-wave solutions, we define a scalar product on the set of the solutions of our Hamilton-Jacobi equations generalized to complex values,
\begin{equation}
\left\langle V_1,V_2\right\rangle = \frac{i}{\hbar}\int \!\!\!d\zeta \, V^\dagger_1(\zeta,\tau)\eta V_2(\zeta,\tau),
\label{eq:ScalarProduct}
\end{equation}
where \(V\) is the eight-dimensional vector 
\begin{equation}
V=\left(A \, P_1 \, P_2 \, P_3 \, \Pi_A \, \Pi_{P_1} \, \Pi_{P_2} \,\Pi_{P_3}\right)^T,
\label{eq:Vdef}
\end{equation}
and \(\eta\) is the selection matrix $\eta = \bigl(\begin{smallmatrix}0&I_4\\ -I_4&0\end{smallmatrix} \bigr)$ with \(I_4\) being the \(4\times 4\) identity matrix.
This scalar product is a conserved quantity in \(\tau\) and therefore the total norm of the quantum states is conserved.

It can also be shown that modes with positive laboratory frequency have positive norm, and those with negative frequency have negative norm \cite{philbin2008,finazzi2013,belgiorno2015}.
In other words, modes belonging to the upper (lower)-half plane in fig.\ref{fig:LabDispRel} have positive (negative) norm.
In the comoving frame, positive frequency waves with negative norm appear, which are associated with spontaneous emission from the quantum vacuum. These generated waves have negative laboratory frequency. Negative norm modes were recently observed in water experiments \cite{weinfurtner2011,rousseaux2008} and in optics \cite{nrr2012,koenig2014}. Due to the conservation of norm, the generation of negative norm waves signifies a simultaneous increase in positive norm waves, the generation of correlated waves.

\section{Mode configurations at a refractive index front}\label{sec:Theorytwo}

In the previous section we presented a canonical model of light and matter interaction in a dielectric medium.
In particular, we highlighted the existence of positive and negative frequency modes.
We now extend this model to a non-uniform medium.\par

The simple geometry of a refractive index front (RIF) is shown in fig. \ref{fig:system} in the comoving frame. The medium is composed of  two homogeneous regions, separated by the RIF at \(\zeta=0\) creating a step in the refractive index.  
The boundary at \(\zeta=0\) constitutes an infinitely steep RIF which propagates in a steady and rigid way in the positive \(\zeta\) direction.
The right (left) region has dispersion parameters \(\kappa_{iR}\)(\(\kappa_{iL}\)) and \(\lambda_{iR}\)(\(\lambda_{iL}\)).
Experimentally, this front could be realized by an asymmetric intense pulse of light with a very steep front on one side, similar to a self-steepened soliton pulse \cite{Agrawal2006}.
The front is approximately infinitely steep if it is shorter than a wavelength.
The step height is the refractive index change between the regions, \(\delta n\), defined by
\begin{equation}
n\left(\zeta\right)=n_L\:\theta \left( -\zeta\right) + n_R\:\theta \left(\zeta\right)=n_R +\delta n\:\theta \left( -\zeta\right)
\label{eq:deltan}
\end{equation}
and illustrated in fig. \ref{fig:system}; $\theta \left(\zeta\right)$ is the Heaviside step function. The index change is described by the scaled Sellmeier coefficients $\beta_{iL}=\sigma \beta_{iR}$ and $\lambda_{iL}^2 = \sigma \lambda_{iR}^2$, where: 
\begin{equation}
\sigma \approx 1+\frac{2 n_R \, \delta n}{n_R^2-1} 
\label{eq:scaleCoeff}
\end{equation}
and $n_R$ is the refractive index on the right side \cite{finazzi2013}. 

In order to identify the allowed modes of propagation, we consider plane wave solutions subject to the dispersion relation in both regions for a given comoving frame frequency \(\omega^{\prime}=\gamma \left(\omega - u k \right)\). This is useful, because the frequency \(\omega^{\prime}\) is conserved, an expression of energy conservation in the moving frame \cite{philbin2008,finazzi2013,belgiorno2015}.
Solutions of fixed \(\omega^{\prime}\) are found at the intersection points between a line of constant  \(\omega^{\prime}\) with the various polariton branches in the dispersion diagram (red circles in fig. \ref{fig:LabDispRel}).
Note that we only consider low positive comoving frequencies \(\omega^{\prime}\) without solutions in the uppermost branch.
At any frequency \(\omega^{\prime}\) there exists a set of 8 discrete \(\left(\omega,k\right)\) solutions to the dispersion relation (\ref{eq:LabSellmeier}). We either find 8 propagating modes or 6 propagating modes and two exponentially growing/decaying modes, that take on complex $\omega$ and $k$.

Emission spectra with 8 propagating modes on only one side of the boundary were calculated in \cite{finazzi2013,finazzi2014}. Here we address the experimentally relevant case allowing for 8 modes to propagate on either side of small refractive index changes.
In particular, consider the positive frequency optical branch (where \(\lambda_2<\lambda<\lambda_3\)) in the moving frame, depicted in fig. \ref{fig:movdisprela}. The black (orange) curve is the branch on the right (left) side of the RIF. Again, the number of solutions depends on  \(\omega^{\prime}\).
On either side of the RIF there is at least one propagating optical mode for all \(\omega^{\prime}\). There also is a frequency interval where three propagating modes exist: between the two horizontal dashed black and the two dashed orange lines in Fig. \ref{fig:movdisprela}, corresponding to the right and left region, respectively. Hereafter these frequency intervals on either side of the RIF are referred to as the \textit{subluminal intervals} (SLI) \(\left[\omega^\prime_{\mathrm{min}L},\omega^\prime_{\mathrm{max}L}\right]\) and \(\left[\omega^\prime_{\mathrm{min}R},\omega^\prime_{\mathrm{max}R}\right]\).\par

Inside a SLI, one of the three mode solutions has a positive comoving group velocity \(\frac{\partial\omega'}{\partial k'}\).
This unique mode allows light on the right of the RIF to escape from it. This middle optical mode (cf. fig. \ref{fig:LabDispRel}) on the right is called \textit{moR} in what follows.
The other two modes have negative comoving group velocity; they move into the boundary from the right. There is a lower (upper) optical mode \textit{loR} (\textit{uoR}). On either side we can order the modes by the comoving wave number $k'$ and obtain \(k_{\omega^{\prime}}^{\prime \mathrm{loR/L}}<k_{\omega^{\prime}}^{\prime \mathrm{moR/L}}<k_{\omega^{\prime}}^{\prime \mathrm{uoR/L}}\) (cf. fig. \ref{fig:movdisprela}).
In the laboratory frame this translates into \(\omega_{\omega^{\prime}}^{\mathrm{loR/L}}<\omega_{\omega^{\prime}}^
{\mathrm{moR/L}}<\omega_{\omega^{\prime}}^{\mathrm{uoR/L}}\)  (fig. \ref{fig:LabDispRel}). All three optical-branch modes of the SLI have positive \textit{laboratory} group velocity on both sides of the boundary: to an observer in the laboratory frame they propagate in the same direction as the RIF.

Beyond this interval (i.e. \(\omega'\notin \left[ \omega^{\prime}_{\mathrm{min}}, \omega^{\prime}_{\mathrm{max}} \right]\)) only one propagating mode remains.
Two complex-wavenumber roots emerge as pairs of exponentially growing and decaying modes that do not propagate.
For \(\omega^{\prime}<\omega^{\prime}_{\mathrm{min}}\) only mode \textit{uoR/L} remains a propagating mode whereas for \(\omega^{\prime}>\omega^{\prime}_{\mathrm{max}}\), only \textit{loR/L}  remains.  It can also be seen in fig. \ref{fig:movdisprela} that for all \(\omega^{\prime}\) there is one propagating mode that belongs to the negative optical-frequencies branch.
This mode has negative norm and will hereafter be referred to as mode \textit{noR/L}.

The subluminal frequency interval of the left region is in general different from that of the right region. Therefore, there exist 5 different combinations of modes across the RIF, also shown in fig. \ref{fig:modestructure}:
\begin{enumerate}
	\item 
\(\omega^{\prime}<\omega^{\prime}_{\mathrm{min}L}\): one optical propagating mode (\textit{uoR/L}) exists (real $\omega$ and \(k\)) on either side of the boundary. All propagating modes exhibit negative group velocities in the moving frame and therefore no optical horizons exist.
	\item \(\omega^{\prime}_{\mathrm{min}L}<\omega^{\prime}<\omega^{\prime}_{\mathrm{min}R}\): three propagating modes (\textit{loL}, \textit{moL}, and \textit{uoL}) on the left of the boundary exist, but only one mode (\textit{uoR}) on the right. Only mode \textit{moL} has positive group velocity on the left. Light in this mode experiences a white hole horizon at the RIF as it can approach, but not enter the right region. To our knowledge, this mode configuration has not yet been described.
	\item \(\omega^{\prime}_{\mathrm{min}R}<\omega^{\prime}<\omega^{\prime}_{\mathrm{max}L}\): three propagating modes (\textit{loR/L}, \textit{moR/L}, and \textit{uoR/L}) exist on either side of the boundary. Modes with negative and positive group velocity exist on either side of the RIF, therefore the RIF is not a one-way door and no horizon exists for waves of this moving frame frequency.
	\item \(\omega^{\prime}_{\mathrm{max}L}<\omega^{\prime}<\omega^{\prime}_{\mathrm{max}R}\): only one propagating mode (\textit{loL}) exists on the left of the boundary, but three modes (\textit{loR}, \textit{moR}, and \textit{uoR}) do on the right. Only mode \textit{moR} has positive group velocity on the right side. Light experiences a black hole horizon at the RIF as it cannot propagate to the right from beyond the RIF. The region on the left (right) of the boundary corresponds to the inner (outer) region of the analogue horizon.
	\item 
\(\omega^{\prime}_{\mathrm{min}R}<\omega^{\prime}\): one propagating mode (\textit{loR/L}) exists on either side of the boundary. All propagating modes exhibit negative group velocities and therefore no optical horizons exist.
\end{enumerate}
Note that for very high refractive indices on the left there is only mode \textit{loL} and no subluminal interval in this region and out of the general case only configurations $1$ (with \textit{loL} instead of \textit{uoL}), $4$, and $5$ remain. 
In this case, modes within the SLI on the right of the boundary always follow configuration $4$.
The 5 configurations introduced here describe a typical experimental combination of horizons with low refractive index changes.\par

\section{Scattering at the RIF}\label{sec:Theoryfour}

Having derived solutions on either side of the RIF, we are now constructing `global' solutions, i.e. solutions to the equation of motion that are valid in both regions. These modes correspond to waves scattering at the RIF and they describe the conversion of an incoming field, even if in the quantum vacuum state, to scattered fields in both regions. We follow the approach \cite{finazzi2013,hopfield1958,barnett1992} to construct these modes and their scattering matrix and then to quantise the solutions to find the photon fluxes  due to spontaneous particle creation. 

The solutions to the dispersion relation correspond to local modes (LMs) that are defined in either the left or right region. We construct `global modes' (GMs) $\mathcal{V}$ as:
\begin{equation}
\mathcal{V} = \sum_\alpha L^\alpha V^\alpha_L \, \theta(-\zeta)+\sum_\alpha R^\alpha V^\alpha_R \, \theta(\zeta),
\label{eq:GMs}
\end{equation}
where $L^\alpha$ ($R^\alpha$) describes the strength of mode $\alpha$ on the left (right) side of the RIF. It can be shown that across the boundary, vector potential, polarisation fields, and their derivatives are continuous \cite{finazzi2013}, which constrains half of the coefficients in Eq. (\ref{eq:GMs}). 
We follow the procedure described in \cite{parentani2009} and consider GMs whose asymptotic decomposition comprises only a single LM with group velocity towards (\textit{in}) or away from (\textit{out}) the RIF.  Thus there are as many of these GMs as there are propagating local modes. Half of the GMs emerge from a defining LM $\alpha$ that moves towards the RIF, forming `global \textit{in} modes' $\mathcal{V}^{\mathrm{in} \alpha}$. The other GMs are `global \textit{out} modes' $\mathcal{V}^{\mathrm{out} \alpha}$, if $\alpha$ is a LM now moving away from the RIF.  The sets of  $\mathcal{V}^{\mathrm{in}}$ and $\mathcal{V}^{\mathrm{out}}$ modes are two basis sets of modes, because the LMs are the complete physical (i.e. non-divergent) solutions in the asymptotic regions.  Hence the scattering matrix $S$ is the transformation of modes from the out-basis to the in-basis:
\begin{equation}
\mathcal{V}^{\mathrm{in},\alpha} = \sum_\beta S^{\alpha \beta} \mathcal{V}^{\mathrm{out},\beta}.
\label{eq:S}
\end{equation}
Scattering and spontaneous photon creation occurs as the input state vacuum does not correspond to the vacuum state in the out-basis, i.e. the spontaneous emission follows from $S$.

We postulate the equivalent of the standard equal-time commutation relations on \(A\) and \(P_i\) and thus quantize the local field  modes  and their momenta:
\begin{equation}
\label{eq:ComRel}
\left[A(\zeta),\Pi_A(\zeta^{\prime})\right]=i \hbar \, \delta\!\left(\zeta-\zeta^{\prime}\right), \quad \left[P_i(\zeta), \Pi_{P_j}(\zeta^{\prime})\right]=i \hbar \, \delta_{ij} \, \delta\! \left(\zeta-\zeta^{\prime}\right).
\end{equation}
We expand the real field \(V\) (Eq. (\ref{eq:Vdef})) on a basis of local frequency eigenmodes
\begin{equation}
V=\int{d\omega^\prime\sum_\alpha{\left(V_{\omega^\prime}^\alpha \, \hat{a}_{\omega^\prime}^\alpha+{V_{\omega^\prime}^{\alpha}}^{\ast} \, \hat{a}_{\omega^\prime}^{\alpha \dagger}\right)}}
\label{eq:Vexpansion}
\end{equation}
that are properly normalized with respect to the scalar product (\ref{eq:ScalarProduct}) under the condition
\begin{equation}
\left|\left\langle V_{\omega_1^\prime}^{\alpha_1},V_{\omega_2^\prime}^{\alpha_2}\right\rangle\right|=\delta\!\left(\omega_2^\prime-\omega_1^\prime\right)\delta_{\alpha_2 \alpha_1}.
\label{eq:NormCond}
\end{equation}
The operators \(\hat{a}_{\omega^\prime}^\alpha\) and \(\hat{a}_{\omega^\prime}^{\alpha \dagger}\) are the annihilation and creation operators of the field mode $\alpha$. Alternatively, we can expand the field over positive frequencies only, including negative norm modes in the expansion:
\begin{equation}
V=\int_0^\infty{d\omega^\prime \left( \sum_{\alpha \in P}{V_{\omega^\prime}^\alpha \, \hat{a}_{\omega^\prime}^\alpha+\sum_{\alpha \in N} V_{\omega^\prime}^{\alpha} \, \hat{a}_{\omega^\prime}^{\alpha \dagger}}\right)}+H.c.,
\label{eq:Vexpansionpos}
\end{equation}
where $P$ ($N$) is the mode set of positive (negative) norms.
Writing the global field $\mathcal{V}$ in the basis of global-in-modes induces quantisation onto the GMs:
\begin{equation}
\mathcal{V}=\int_0^\infty{d\omega^\prime \left( \sum_{\alpha \in P}{\mathcal{V}_{\omega^\prime}^{\mathrm{in},\alpha }\, \hat{a}_{\omega^\prime}^{\mathrm{in},\alpha}+\sum_{\alpha \in N} \mathcal{V}_{\omega^\prime}^{\mathrm{in}, \alpha} \, \hat{a}_{\omega^\prime}^{\mathrm{in},\alpha \dagger}}\right)}+H.c.
\label{eq:GVexpansion}
\end{equation}
and a similar expression can be written for $\mathcal{V}$ using the global-out-modes.  The expansion (\ref{eq:GVexpansion})  for in- and out modes defines the annihilation and creation operators for the global modes as well as the transformation between in- and out creation and annihilation operators. If $\hat{A}^{\mathrm{in}}$ is a row vector containing all the annihilation (or creation) operators for positive (or negative) norm global-in-modes and $\hat{A}^{\mathrm{out}}$ correspondingly for the out-modes, then the transformation of operators is \cite{finazzi2013}:
\begin{equation}
\hat{A}^{\mathrm{out}}=\hat{A}^{\mathrm{in}} \, S.
\label{eq:FockTransform}
\end{equation}
Using Eq. (\ref{eq:FockTransform}) we obtain the flux density of photons $I'^\alpha$ in mode $\alpha$, the number of particles per unit time $\Delta \tau$ and bandwidth in the moving frame, as \cite{finazzi2013}:
\begin{equation}
I'^\alpha= \frac{2 \pi \left\langle 0 |\hat{a}_{\omega^\prime}^{\alpha \dagger} \hat{a}_{\omega^\prime}^\alpha|0\right\rangle }{\Delta \tau}.
\label{eq:Ip}
\end{equation}
In terms of the scattering matrix, the flux density in mode $\alpha$ follows from Eq. (\ref{eq:FockTransform}) and Eq. (\ref{eq:Ip}):
\begin{equation}
I'^\alpha=\sum_{\bar{\alpha}} \left|S^{\bar{\alpha} \alpha}\right|^2,
\label{eq:IpScatt}
\end{equation}
where $\bar{\alpha}$ is a mode with a norm opposite in sign to $\alpha$. Spectra for positive norm modes and for single SLIs were calculated with this relation in \cite{finazzi2013,finazzi2014}. We now move on to obtain the spectra of emission for any frequency in all modes, all mode configurations (see sec. \ref{sec:Theorytwo}), and for a variety of refractive index changes $\delta n$.

\section{Emission spectra and photon flux}\label{sec:Results}

We proceed to use the S matrix to compute the spectra of emission into all optical modes as seen from the moving and the laboratory frame.
We consider modes in bulk fused silica.  The material resonances are \(\lambda_{1,2,3}=9,904\,\)nm, \(116\,\)nm and \(68.5\,\)nm, respectively, and the elastic constants are \(\kappa_{1,2,3}=0.07142,0.03246\), and 0.05540, respectively. The velocity of the RIF is \(u=0.66\, c\), corresponding to a group index of $1.5$. 

We first consider spectra in the moving frame. Fig. \ref{fig:Cascade} displays the spectrum of emission into \textit{moR}, the unique rightgoing mode (black hole emission), for different index changes $\delta n$. Spectral emission is constrained to the right SLI $\left[ \omega^\prime_{\mathrm{min}R},\omega^\prime_{\mathrm{max}R} \right]$, where the mode \textit{moR} exists. An optical horizon, however, does only exist for part of this interval, i.e.  \( \left[ \omega^\prime_{\mathrm{max}L},\omega^\prime_{\mathrm{max}R} \right] \), because at lower frequencies the SLIs of the left and right overlap (cf. fig. \ref{fig:movdisprela}).  As can be clearly seen, the absence of a horizon leads to a significant decrease in the emission, i.e. mode coupling, although some emission remains.  In case of at a large index change (e.g. $\delta n = 0.056$), as calculated and observed in \cite{finazzi2013}, the spectral shape is not thermal as is expected from a finite subluminal interval. Here we observe a remarkable feature: the shape of the spectrum above $\omega^\prime_{\mathrm{max}L}$, where a horizon exists, is independent of the index change. This can be more clearly seen in fig. \ref{fig:shapematch}, in which the spectra for lower index changes are scaled up to compensate the lower single frequency rate. All traces over orders of magnitude of index changes line up to the same shape, making it a universal signature of analogue black hole emission. Note also that the shape differs for emission outside the SLI.

Next we consider spectra of emission into each mode (i.e. \textit{noL}, \textit{loL}, \textit{uoL}, \textit{moR}) for a fixed $\delta n =0.02$, as displayed in fig. \ref{fig:allopticalmodes}. The strongest emission occurs into the optical mode with negative norm, \textit{noL}. This emission is due to coupling with all the other positive norm modes in the medium and is strongest where this superluminal mode couples to a subluminal mode ($ \omega^\prime_{\mathrm{max}L}<\omega^\prime<\omega^\prime_{\mathrm{max}R}$). 

It is interesting to calculate the total photon flux over the SLI \( \left[ \omega^\prime_{\mathrm{min}R},\omega^\prime_{\mathrm{max}R} \right] \) by integrating over the spectrum of fig. \ref{fig:Cascade}. In order to convert the flux to a realistic, though very approximate photon number, we assume that the RIF propagates over a distance of $1\,$mm.  The resulting photon number as a function of index change $\delta n$ is given in fig. \ref{fig:blackholeinterval}a).
The number of photons excited from the vacuum first grows with power $\approx 2.5$ of \(\delta n\) until \(\delta n=0.052\). The emission spectrum becomes wider in a linear way, as shown in  fig. \ref{fig:blackholeinterval}b). Thus the emission rate for a single mode increases with \(\delta n^{3/2}\).  This result is surprising and will be further investigated elsewhere. 
For very high induced index change, i.e. for \(\delta n\geq 0.052\), the spectral width saturates and the emission rate grows slower accordingly. These index steps are difficult to reach experimentally by nonlinear optical pulses though.

The spectra calculated above would be observable in the moving frame. In an actual experiment the detectors are located in the laboratory frame and thus the spectrum observed is different. For each mode, the rate of photon production per unit time and unit frequency \(\omega\) in the laboratory frame is derived from the moving frame rate as \cite{finazzi2013}:
\begin{equation}
I^\alpha=\left( 1- \frac{u }{v_g(\omega)}\right) I'^\alpha ,
\label{eq:I}
\end{equation}
where $v_g(\omega)$ is the laboratory group velocity at $\omega$. The spectral density as a function of frequency converts to wavelength by the factor \({\omega^2}/({2 \pi c})\). 
A lab spectrum for a single mode was calculated in \cite{finazzi2013}. However, this spectrum is not observable for the following reason: on  either side of the RIF, each moving frame frequency $\omega'$ corresponds to up to eight different laboratory frequencies depending on the mode involved, as in fig. \ref{fig:LabDispRel}. Note that we calculate the emission with positive laboratory group velocity only. Thus each mode has a unique direction of group velocity in both frames independent of the side of the RIF. Hence, of the modes left and right of the RIF, half are defining global out (emission) modes. Conversely, the emission at a fixed laboratory frequency may arise from several optical modes.
We proceed to calculate for successive frequencies $\omega$ the emission rates in each associated global out mode, obtaining a laboratory frame spectrum for each mode. We then add up the spectra to obtain the total spectral emission.
Starting at $\omega_m$,  the lowest positive optical frequency, we obtain emission into outgoing mode \textit{loL} on the left, then \textit{moR} on the right followed by \textit{uoL} on the left and finally  \textit{noL} on the left.

Fig. \ref{fig:LSDlambda} shows laboratory spectra for three index changes. 
As the spectrum is composed of contributions from different modes for different mode configurations, it exhibits a number of sharp features. The largest spectral density is obtained around $250\,$nm and corresponds to emission from the negative norm mode \textit{noL}. Emission is generated by the pairwise coupling of two modes of opposite norm. Hence mode \textit{noL} is the only negative norm mode on the optical branch and covers a rather small laboratory spectral interval, between the violet and red lines in fig. \ref{fig:LSDlambda}. Therefore, all emission due to the coupling of two optical modes leaves a contribution within this emission peak in the UV spectral range. The coupled positive norm mode (i.e. the partner photon), if optical, can be found at the remaining optical frequencies. We choose to limit the range of optical wavelengths such that no modes in the uppermost branch are excited, resulting in a cut off at $230\,$nm.  Not all coupled mode pairs are separated by a black or white hole-type horizon. For example, intervals with horizons, as schematically sketched in fig. \ref{fig:modestructure}, are found between the black and orange dashed lines in fig.  \ref{fig:LSDlambda}, but not in the adjacent spectral regions. The short (long) wavelength interval  corresponds to a black (white) hole configuration. 
The presence of optical horizons leads to an enhancement of the emission on the background of the horizonless emission that seems to decay from the ultraviolet to the infrared. 
Modes \textit{moR} and \textit{loL} exhibit clear horizon emission profiles between the black and orange dashed lines and their intervals of emssion are indicated by arrows. Over the visible range,  emission from \textit{moR} dominates. Fig. \ref{fig:LSDlambda} also shows traces for lower nonlinearities. As expected, the spectral density decreases and the invervals of optical horizons, associated with strong emission, become smaller. 
The red line corresponds to zero moving frame frequency $\omega'$; no major spectral features seem to be associated with this position.

The quantum state at the output is expected to be a two-mode squeezed vacuum state, if two modes only were involved. However, for each moving frame frequency, each mode can couple to up to five positive norm and three negative norm modes. Thus, we expect the final quantum state on the optical branch to be in a partly mixed state across the optical modes. Yet, coupling between particular mode pairs seems to dominate in parts of the spectrum, in particular within the optical branch. Further characterization of the exact state emerging is needed.

\section{Conclusion}\label{sec:conclusion}
Emission of light from the quantum vacuum has recently attracted attention \cite{philbin2008,finazzi2013,finazzi2014,belgiorno2014} due to the high emission rates expected in the presence of an analogue horizon and the hope to observe analogue Hawking radiation. Using a fully analytical model, introduced in \cite{finazzi2013}, the spontaneous emission of light from a moving dielectric boundary, a refractive index front (RIF), can be calculated. The model can be used to study any dispersive medium phenomenologically described by Sellmeier equations.

In this work we have focussed on the case of SLIs on both sides of the RIF.  We calculate the emission for a system based on bulk fused silica.  Our approach is interesting, because we can show how the emission is structured into intervals with black hole-, white hole, and no horizon and how these horizons lead to an increase in the spectral density in the laboratory over a general decay.  To our understanding, the enhancement of emission due to horizons is a common feature of realistic refractive index changes.  For index changes beyond the damage threshold no enhancement was observed [19].  Additionally, because we consider contributions from all modes, we have found that the emission is dominated and characterized by a peak in the (laboratory) UV that corresponds to photons emitted in the negative norm mode.  We show that the emission spectrum from the analogue black hole has a unique shape, which is independent of the index change.  It is not of the thermal shape that is expected from a nondispersive analogue black hole horizon.  We also calculate the emitted flux and, as a result, the flux increases as a power law of exponent 2.5 for lower index steps, as more and more frequency modes appear to participate in the emission at the analogue black hole horizon.  Whereas previous calculations have mainly focused on the Hawking-like emission in the moving frame, these results reveal signatures of the quantum vacuum emission important for the experimental observation, such as the nonlinear scaling, spectral shape, or mode configuration.  

This framework forms the basis to future calculations of the revealing quantum correlations between different outgoing modes, correlations of Hawking partners as well as horizonless emission into other modes. This analysis would allow for a better prediction of the exact quantum state produced, a key connection to the Hawking effect.  The analysis can also be extended to more general variations of the refractive index as well as broader choices of media, such as waveguides.

\section{Acknowledgements}\label{sec:acknowledgements}

The authors would like to acknowledge useful discussions with S. Finazzi and R. Parentani, as well as support from EPSRC via grant EP/L505079/1.

\setcounter{figure}{0}

\begin{figure}[H]
	\centering
		\includegraphics[scale=1]{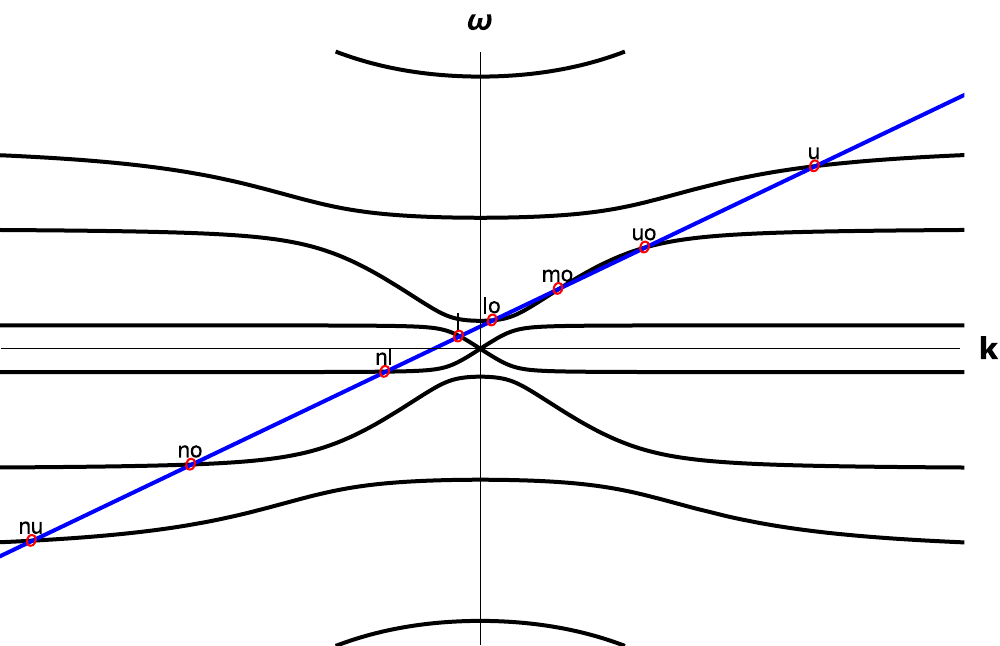}
		
	\caption{Sellmeier dispersion relation, Eq. (\ref{eq:LabSellmeier}), with three resonances in the laboratory frame. There are eight branches (black lines). A contour of  $\omega^{\prime}$ is shown in blue. Their intersection points indicate the modes of propagation in the medium (red).}
	\label{fig:LabDispRel}
\end{figure}

\begin{figure}[H]
	\centering
		\includegraphics[scale=0.80]{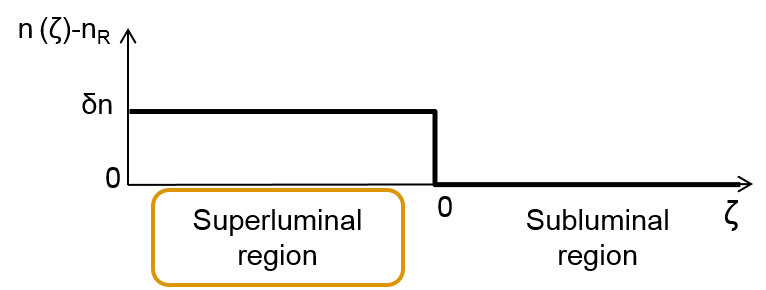}
		
	\caption{Sketch of the refractive index front (RIF) in the moving frame: there are two homogeneous regions of uniform refractive index on the left and right of a dielectric boundary of height \(\delta\)n. For suitable moving frame frequencies, light on the left (right) of the RIF may not (may) propagate away to the right as the step moves at superluminal (subluminal) speeds. Light on the left is trapped behind an analogue horizon at $\zeta = 0$. }
	\label{fig:system}
\end{figure}

\begin{figure}[H]
	\centering
		\includegraphics[scale=0.60]{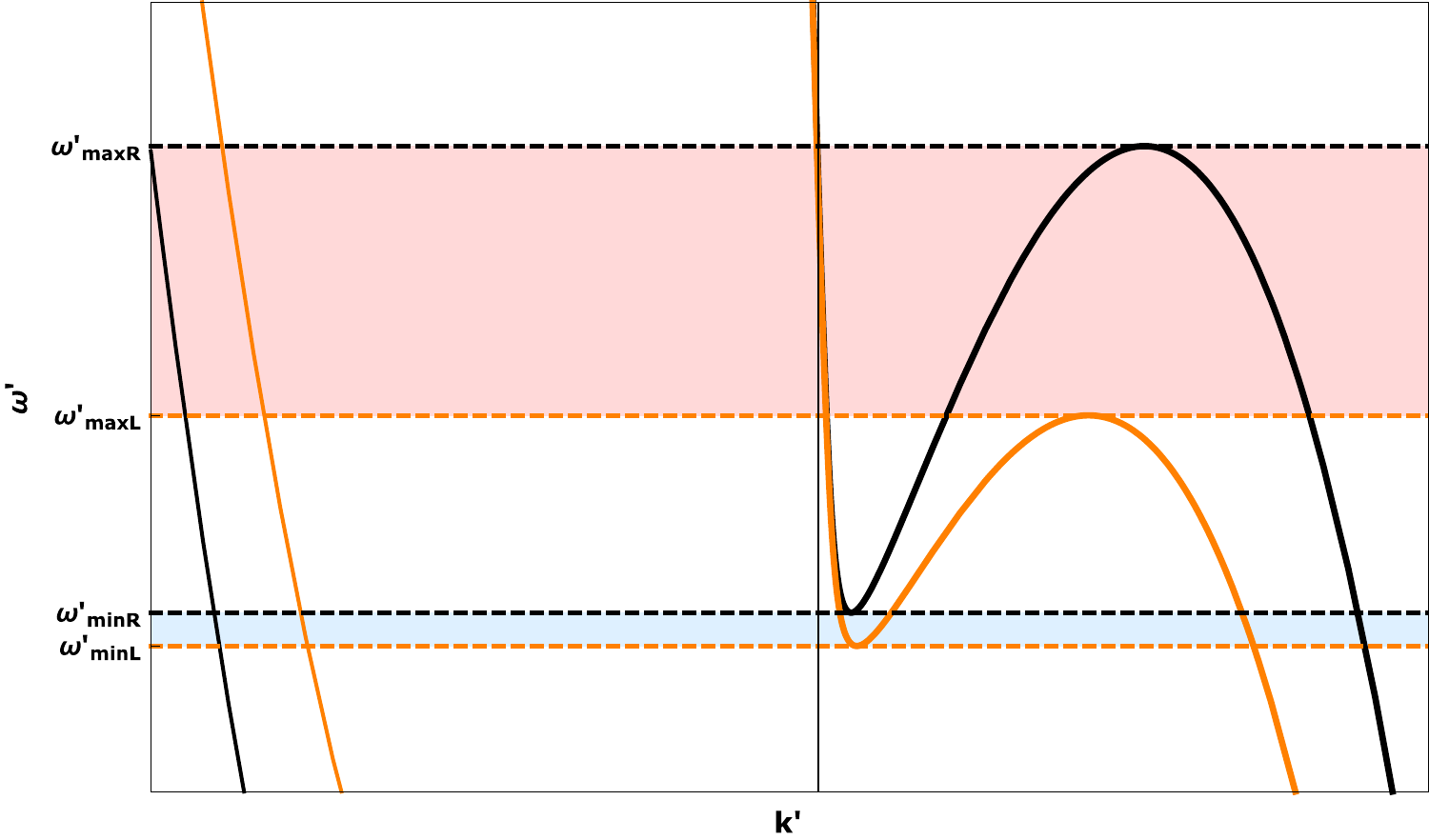}
		
	\caption{Sellmeier dispersion relation of fused silica in a frame moving at a velocity $u=0.66 \, c$.  Part of the optical branch is shown: branches with positive (negative) laboratory frequencies are represented by thick (thin) curves. A curve for zero refractive index change  \(\delta n\) is shown in black and for a high change,  \(\delta n=0.02\), in orange. Frequency intervals corresponding to black and white hole analogue horizons are shaded in orange and blue, respectively.}
	\label{fig:movdisprela}
\end{figure}

\begin{figure}[H]
	\centering
		\includegraphics[scale=1]{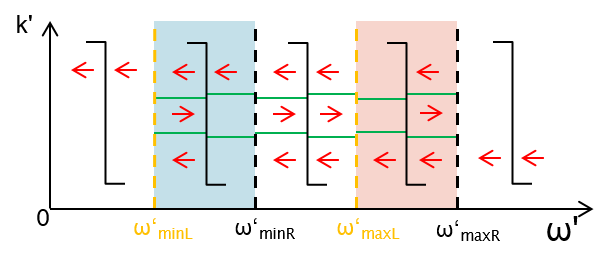}
		
	\caption{Diagrammatic explanation of the possible mode configurations for positive-frequency optical modes for various comoving frequencies. Configurations change if two of the modes share their group velocity with the RIF (green lines), which occurs at the frequencies indicated above and in fig. \ref{fig:movdisprela}.  }
	\label{fig:modestructure}
\end{figure}

\begin{figure}[H]
	\centering
		\includegraphics[scale=1]{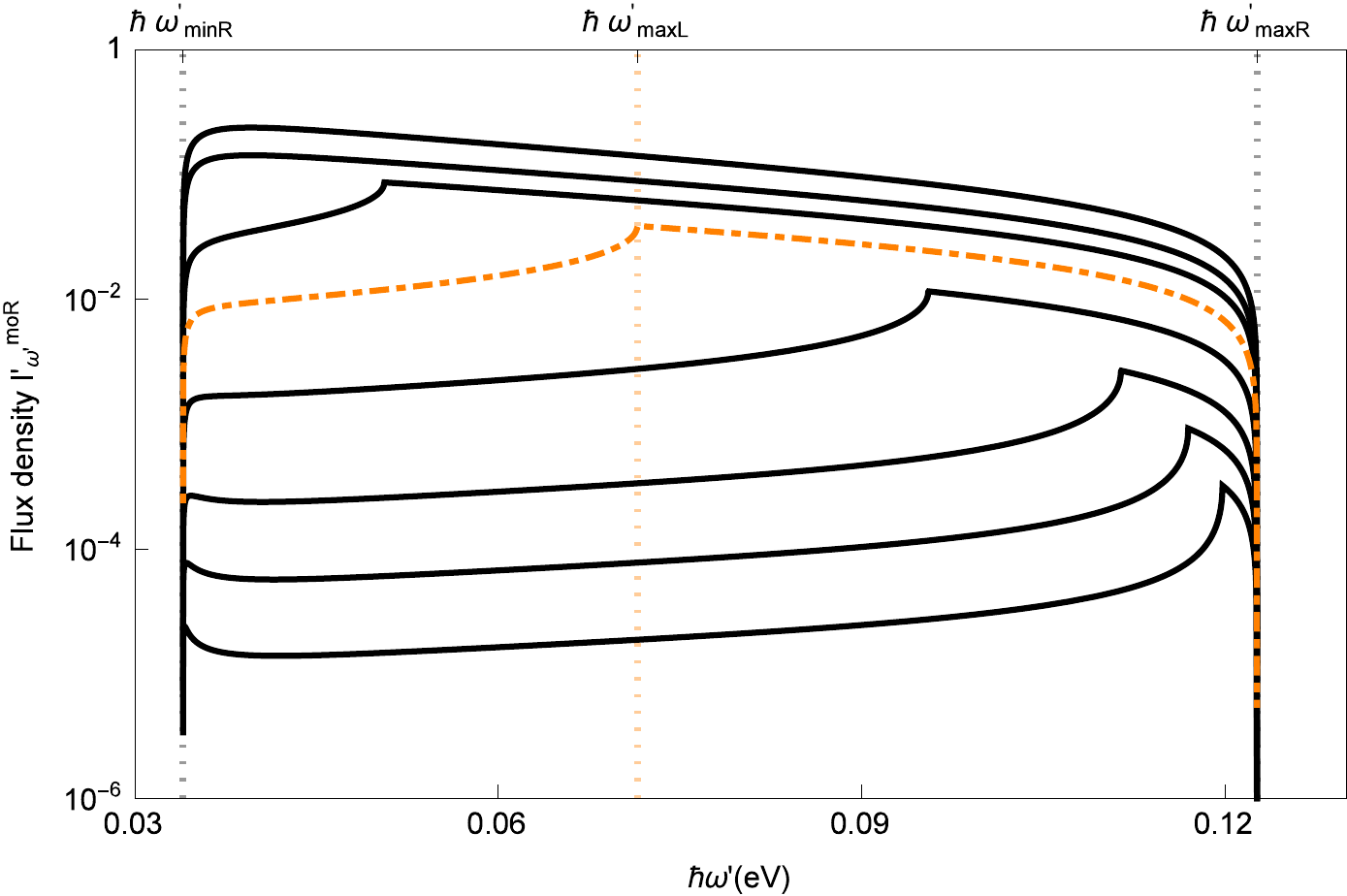}
		
	\caption{Spectrum of emission into the uniquely right propagating mode \textit{moR} on the right side of the RIF. Emission is calculated in the moving frame of velocity $u = 0.66\,c$. The number of particles per time and bandwidth, the flux density, is displayed for increasing values of \(\delta n\) (\(1\ast 10^{-3}\), \(2\ast 10^{-3}\), \(4\ast 10^{-3}\), \(1\ast 10^{-2}\), \(2\ast 10^{-2}\) (orange dot dashed), \(3\ast 10^{-2}\), \(4\ast 10^{-2}\), \(5.6\ast 10^{-2}\)). The spectrum is confined to a finite interval over which the mode \textit{moR} exists.}
	\label{fig:Cascade}
\end{figure}

\begin{figure}[H]
	\centering
		\includegraphics[scale=1]{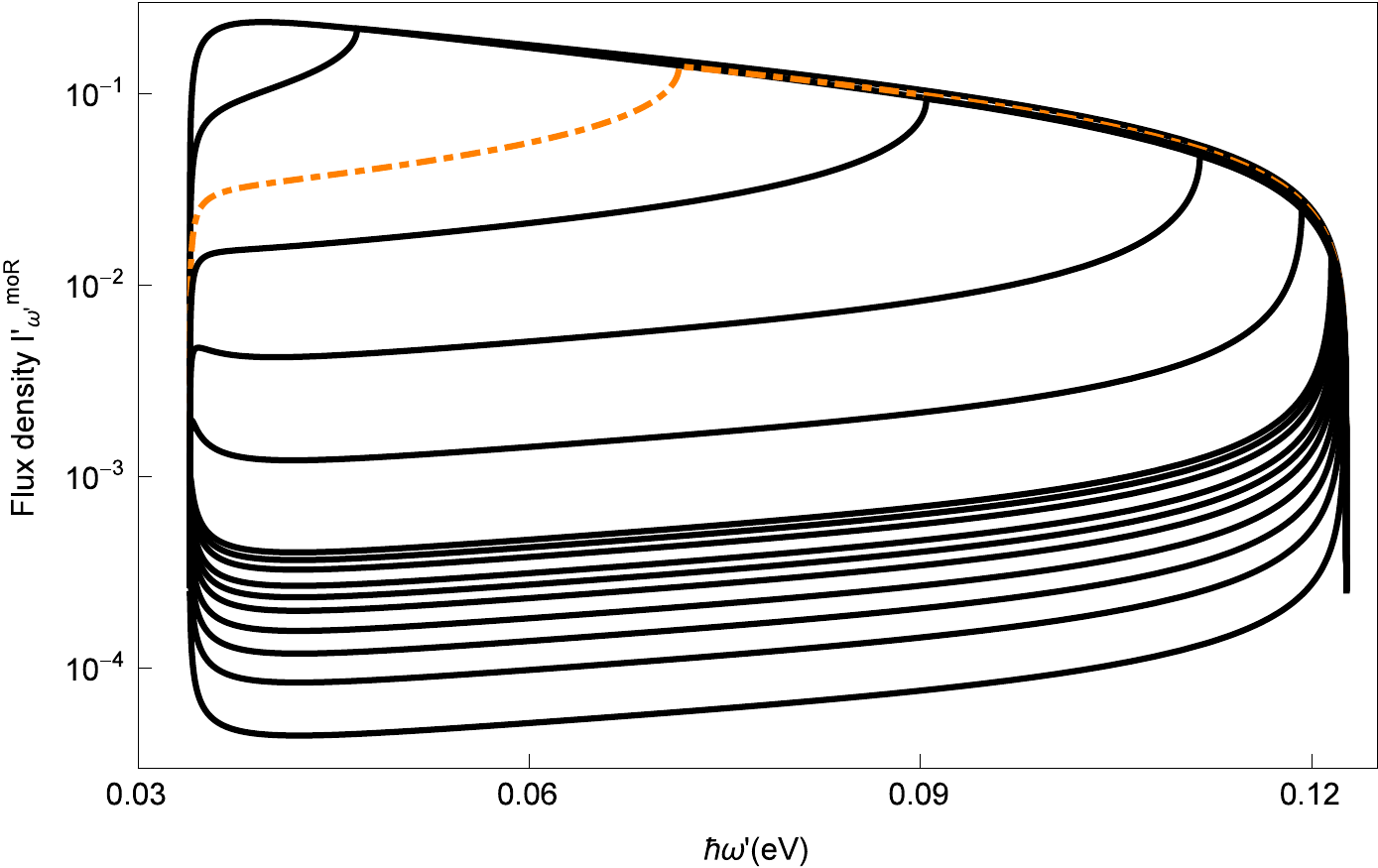}
		
	\caption{Spectrum of emission as in fig. \ref{fig:Cascade}. To compare the shapes of the traces, spectral densities are scaled such that all traces line up with the  \(\delta n=0.02\) trace (orange). Values of \(\delta n\)  range from \(4\ast 10^{-5}\) to  \(5.2\ast 10^{-2}\).}
	\label{fig:shapematch}
\end{figure}

\begin{figure}[H]
	\centering
		\includegraphics[scale=1]{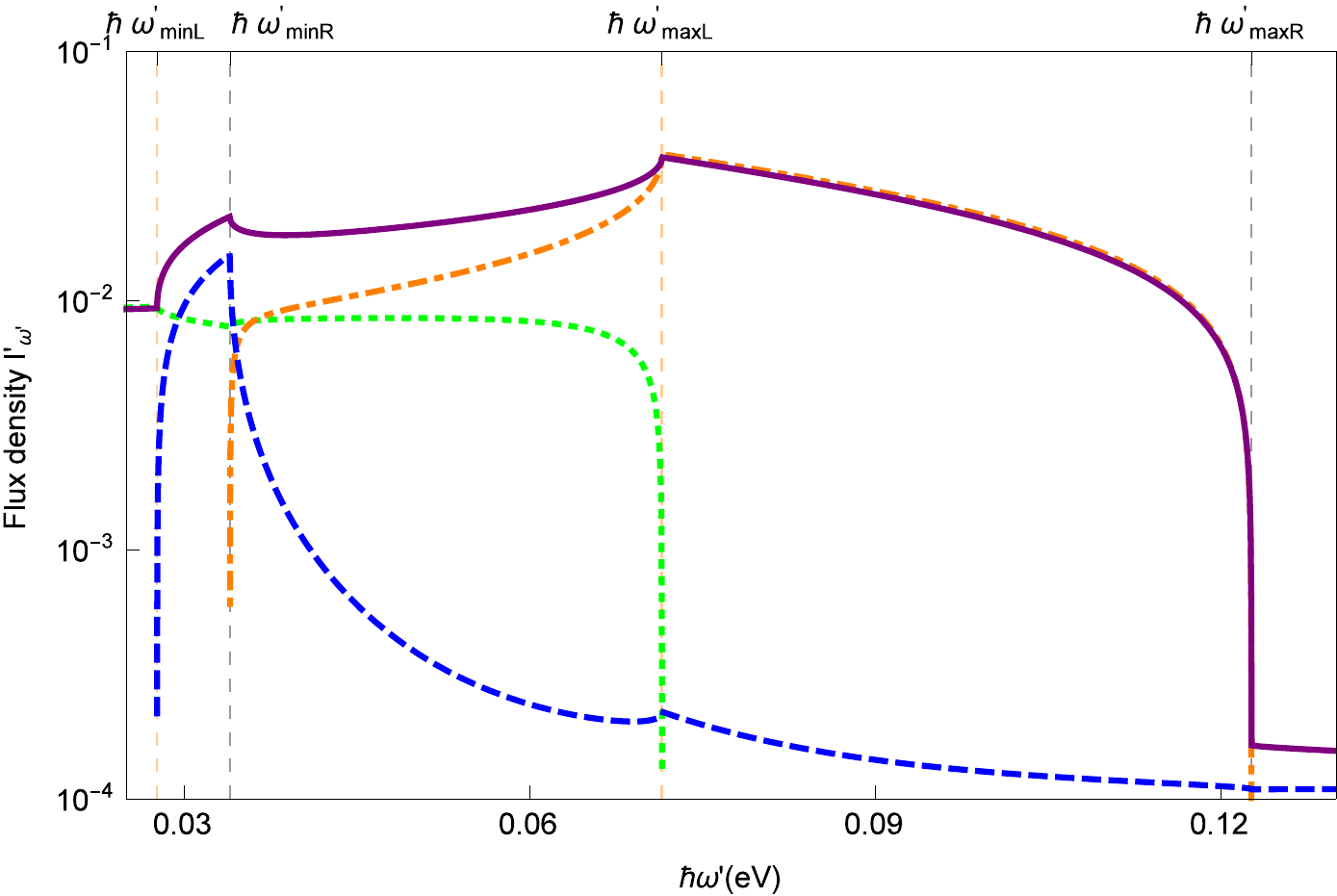}
		
	\caption{Emission spectra of each optical mode in the comoving frame (\(u=0.66\,c\),\,\(\delta n=0.02\)).  Emission into mode \textit{noL}: purple;  \textit{loL}: blue dashed; \textit{uoL}: green dotted; \textit{moR}: orange dot dashed.}
	\label{fig:allopticalmodes}
\end{figure}

\begin{figure}[H]
	\centering
		\includegraphics[scale=1.0]{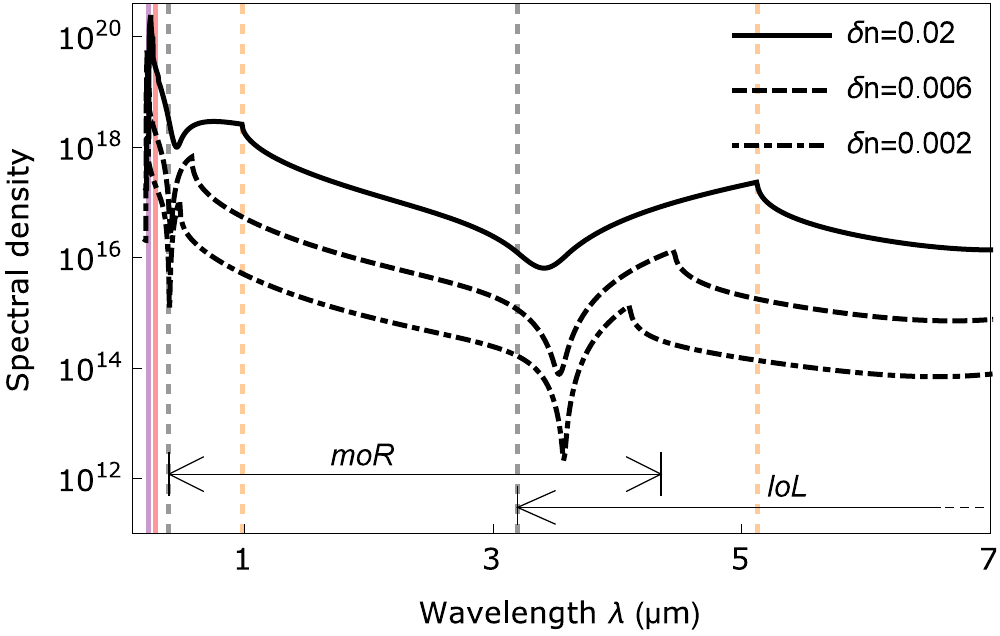}
		
	\caption{Emission spectral density in the laboratory frame.  At each wavelength the total spectral flux density, the number of photons emitted per unit time and unit bandwidth, is the sum of contributions from all modes. Emission is concentrated in the UV in a narrow spectral peak generated from mode \textit{noR}. Outside the peak the emission decreases except for specral intervals corresponding to black- and white- hole analogue horizons.  Spectra are calculated for wavelengths above the violet line, beyond which there are no contributions from the uppermost dispersion branch.  The red line corresponds to $\omega' = 0$ (phase velocity horizon). The black and orange dashed lines indicate the interval of the black (white) hole mode configuration for the \textit{moR} (\textit{loL}) mode at short (long) wavelengths.}
	\label{fig:LSDlambda}
\end{figure}

\begin{figure}
        \centering
        \begin{subfigure}[H]{0.5\textwidth}
                \includegraphics[width=\textwidth]{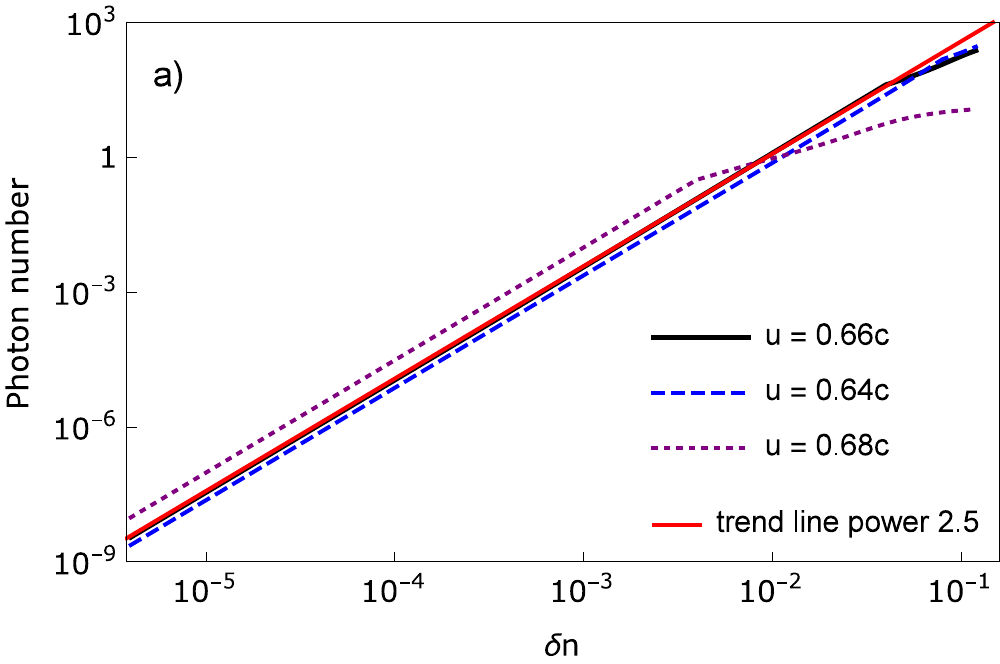}
                \label{fig:photonnumber}
        \end{subfigure}~
             \begin{subfigure}[H]{0.5\textwidth}
                \includegraphics[width=\textwidth]{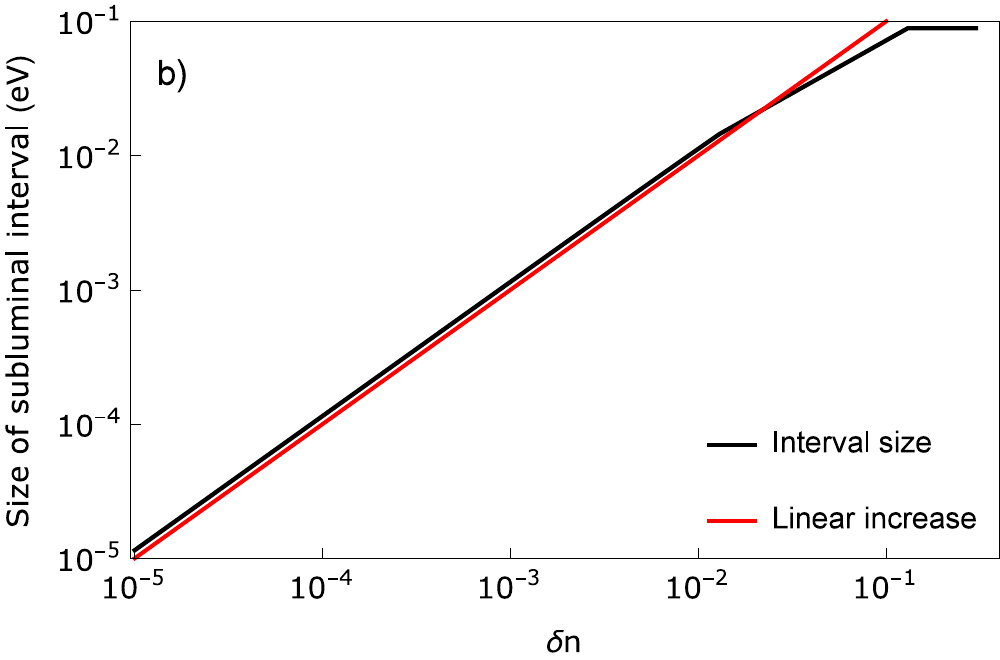}
                \label{fig:sizeinterval}
				\end{subfigure}
				
        \caption{Total emission over the subluminal (black-hole) interval into the unique right moving mode \textit{moR}. Estimated photon number for different velocities $u$ (left panel) and size of the interval in the moving frame (right panel) as a function of index change $\delta n$. }\label{fig:blackholeinterval}
\end{figure}

\end{document}